\documentclass[12pt]{article}

\usepackage[tbtags]{amsmath}
\usepackage{amssymb}
\usepackage{amsfonts}
\usepackage{euscript}
\usepackage{longtable}
\usepackage{graphicx}
\usepackage{color}
\usepackage{shadow}

\begin{document}

\title{\bf Describing orbit space of global unitary actions for mixed qudit states}

\author{Vladimir  Gerdt\, ${}^{a}$\,,  Arsen  Khvedelidze\,${}^{a, b, c}$\,  and  Yuri  Palii \,${}^{a,d}$
\\[1cm]
${}^a$\,\small\it Laboratory of  Information Technologies,  Joint Institute for Nuclear Research, Dubna, Russia\\
${}^b$\,\small\it Iv. Javakhishvili Tbilisi State University,  A.Razmadze Mathematical Institute,  Georgia \\
${}^c$\,\small\it School of Natural Sciences, University of Georgia, Tbilisi, Georgia  \\
${}^d$\,\small\it Institute of Applied Physics, Moldova Academy of Sciences,
Chisinau, Republic of Moldova
}

\date{\empty }

\maketitle

\begin{abstract}
The unitary $ \mathrm{U}(d)\--$equivalence relation on the space $\mathfrak{P}_+\,$  of mixed states  of $d$\--dimensional quantum system defines the orbit space  $ \mathfrak{P}_+/ \mathrm{U}(d)\,$ and provides its description in  terms the ring $\mathbb{R}[\mathfrak{P}_+]^{\mathrm{U}(d)}\,$ of $\mathrm{U}(d)$\--invariant polynomials.
We prove that the semi-algebraic structure of  $ \mathfrak{P}_+/ \mathrm{U}(d)\, $
is determined completely  by  two basic properties of density matrices, their semi-positivity and Hermicity.
Particularly, it is shown that the Procesi-Schwarz inequalities in elements of integrity basis for  $\mathbb{R}[\mathfrak{P}_+]^{\mathrm{U}(d)}\,$  defining the orbit space,   are identically satisfied for all elements of  $\mathfrak{P}_+$.
\end{abstract}

\newpage

\tableofcontents

\newpage

\section{Introduction}

The basic symmetry of isolated quantum systems is the unitary invariance. It sets the equivalence relations between the states and defines the physically relevant factor space.  For composite systems implementation of this symmetry has very specific features leading to a such non-trivial phenomenon as the entanglement of quantum states.

The space of mixed states, $\mathfrak{P}_+\,,$ of  $d$\--dimensional  binary quantum system
is locus in quo  for  two unitary  groups action:  the group  $\mathrm{U}(d)$ and the tensor product group  $\mathrm{U}(d_1)\otimes\mathrm{U}(d_2)\,, $ where $d_1, d_2$ stand for dimensions of subsystems, $d=d_1d_2$.
Both groups act  on a state $\varrho \in \mathfrak{P}_+\,$ in adjoint manner
\begin{equation}
\label{eq:adaction}
(\mathrm{Ad}\,g\,)\varrho  =  g\, \varrho\, g^{-1}\,.
\end{equation}
As a result of this action one can consider two equivalent classes of $\varrho$;  the  \textit{global}
$\mathrm{U}(d)\--$orbit and  the \textit{local }$ \mathrm{U}(d_1)\otimes\mathrm{U}(d_2)\--$orbit.
The collection  of all $\mathrm{U}(d)$\--orbits, together  with the quotient topology and differentiable structure defines the ``global orbit space'',  $\mathfrak{P}_+/ \mathrm{U}(d)\,,$  while the orbit space $\mathfrak{P}_+/  \mathrm{U}(d_1)\otimes\mathrm{U}(d_2)$\,  represents
the ``local orbit space'' , or the so-called \textit{entanglement space} $\mathcal{E}_{d_1\times d_2}$. The latter space is proscenium for manifestations of the intriguing effects occurring in quantum information processing and communications.

Both orbit spaces admit representations  in terms of  the elements of integrity basis  for the corresponding ring of $\mathrm{G}$\--invariant polynomials, where $\mathrm{G}$ is either $\mathrm{G}=\mathrm{U}(d)$ or $ \mathrm{G}=\mathrm{U}(d_1)\otimes\mathrm{U}(d_2)$.   This can be done implementing
the  Procesi and Schwarz method,  introduced in 80th of last century  for description of
the orbit space of a compact Lie group action on a linear space
\cite{ProcesiSchwarz1985PHYSLETT,ProcesiSchwarz1985}.
According to the Procesi and Schwarz the orbit space  is identified with the semi-algebraic variety, defined  by the syzygy ideal  for the integrity basis and the semi-positivity condition of  a special, so-called ``gradient matrix'', $\mathrm{Grad}(z) \geqslant 0\,,$
that is constructed from the integrity basis elements.
In the present note we address the question of application of this generic approach to the construction of
 $\mathfrak{P}_+\,/\mathrm{U}(d)\,$ and $\mathfrak{P}_+/  \mathrm{U}(d_1)\otimes\mathrm{U}(d_2)$.
Namely, we  study whether the semi-positivity of  $\mathrm{Grad}\--$ matrix introduces new conditions on the elements of the integrity basis for the corresponding  ring $\mathbb{R}[\mathfrak{P}_+]^{\mathrm{G}}$.
Below it will be shown that for the global unitary invariance, $\mathrm{G}=\mathrm{U}(d)\,,$ the semi-algebraic structure  of the orbit space is determined solely from the physical conditions on density matrices, their semi-positivity and Hermicity. The conditions $\mathrm{Grad}(z) \geqslant 0\,$ do not bring new restrictions on the elements of integrity basis for  $\mathbb{R}[\mathfrak{P}_+]^{\mathrm{U}(d)}\,.$ Opposite to this case, for the local symmetries the
Procesi and Schwarz inequalities impact on the algebraic and geometric properties of the entanglement space.

Our presentation is organized as follows. In section \ref{sec:Proc-Schwarz} the Procesi and Schwarz method  is briefly
stated in the form applicable to analysis of  adjoint unitary action on the space of states.
In section \ref{sec:P/U} the semi-algebraic structure of $ \mathfrak{P}_+/ \mathrm{U}(d)\, $ is discussed.
The final section is devoted to a detailed consideration of  two examples,
the orbit space of qutrit (d=3) and the global orbit space
of four-level quantum system (d=4).

\section{The Procesi-Schwarz method}
\label{sec:Proc-Schwarz}

Here we briefly state the above mentioned method  for  the orbit space construction  elaborated by Procesi  and Schwartz for the case of compact Lie group
action on a linear space \cite{ProcesiSchwarz1985PHYSLETT,ProcesiSchwarz1985}.

Consider a compact Lie group G acting  linearly  on the real $d$\--dimensional vector space $V$.
Let $\mathbb{R}[V]^{\mathrm{G}}$ is the corresponding ring of the $\mathrm{G}$\--invariant polynomials on $V$.
Assume  ${\cal P} = \left(p_1, p_2, \dots ,p_q\right)$  is a  set of homogeneous polynomials  that form the  integrity basis,
$$\mathbb{R}[x_1, x_2,\dots, x_d]^{\mathrm{G}}= \mathbb{R}[p_1, p_2, \dots , p_q]\,.$$
Elements of the integrity basis define the polynomial mapping:
  \begin{equation}
  \label{eq:polmap}
p:  \qquad V \rightarrow\mathbb{R}^q\, ; \qquad  (x_1, x_2, \dots , x_d)  \rightarrow (p_1, p_2, \dots , p_q)\,.
\end{equation}
Since  $p$ is constant on the orbits of $\mathrm{G}$ it
 induces a homeomorphism of the orbit space $V/G$  and the image $X$ of  $p$\--mapping;  $V/G\simeq X$ \cite{ProcesiSchwarz1985PHYSLETT,ProcesiSchwarz1985}.
In order to describe $X$  in terms of ${\cal P}$  uniquely,  it is necessary to take into  account  the
\textit{syzygy ideal } of  ${\cal P},$  i.e.,
$$
 I_{\cal P}=\{h \in \mathbb{R}[   y_1, y_2, \dots , y_q]\ :\  h(p_1, p_2, \dots, p_q) =0\,\}\subseteq   \mathbb{R}[V ]\,.
$$
Let  $Z \subseteq  \mathbb{R}^q$  denote the  locus of common zeros of all elements of  $I_{\cal P}\,,$  then $Z$ is affine variety in $\mathbb{R}^q\,$ such that $X \subseteq Z$. Denote by $\mathbb{R}[Z]$ the {\em coordinate ring} of $Z$, that is, the ring of polynomial functions on $Z$. Then the following isomorphism
takes place \cite{CoxLittleO'Shea}
$$
\mathbb{R}[Z]\simeq \mathbb{R}[y_1, y_2, \dots , y_q]/I_{\cal P}\simeq \mathbb{R}[V]^{\mathrm{G}}\,.
$$
Therefore, the subset  $Z$ essentially is determined by $\mathbb{R}[V]^{\mathrm{G}}$,
but to describe  $X $ the further steps are required.
According to  \cite{ProcesiSchwarz1985PHYSLETT,ProcesiSchwarz1985} the necessary information on $X$ is encoded in the semi-positivity
of $q\times q $ matrix  with  elements  given by  the  inner products of
gradients, $\mathrm{grad}( p_i ):$
\[ ||\mathrm{Grad}||_{ij}= \left(\mathrm{grad}\left(  p_i \right), \mathrm{grad}\left(  p_j \right)\right)\,.\]

Briefly summarizing all above, the G-orbit space  can be  identified with
the semi-algebraic variety, defined as points,  satisfying two conditions:
\begin{itemize}
\item[a)]
$z\in Z $,  where $Z$ is the surface defined by the syzygy ideal
for the integrity basis
of  $\mathbb{R}[V]^{\mathrm{G}}$;
\item[b)] $\mathrm{Grad}(z) \geqslant 0\,.$
\end{itemize}

Having in mind these basic facts one can pass to the construction  of  the orbit space  $ \mathfrak{P}_+/ \mathrm{U}(d)\, .$  At first  we describe the generic semi-algebraic structure and further exemplify  it considering   two simple, three and four level quantum systems.

\section{Semi-algebraic structure  of
$ \mathfrak{P}_+/\mathrm{U}(d)\, $}
\label{sec:P/U}

The first step making  the Procesi-Schwarz method applicable  to the case we are interested in consists  in the  linearization of the adjoint $\mathrm{U}(d)\--$action
(\ref{eq:adaction}).  For the unitary action one can achieve this as follows.
Consider the space $\mathcal{H}_{d\times d}\, $ of  $d\times d$  Hermitian  matrices and define the mapping
\[
\mathcal{H}_{d\times d} \to \mathbb{R}^{d^2}\,;  \qquad   \varrho_{11}=v_1, \varrho_{12}=v_2, \dots ,\varrho_{1d} =v_{d}, \varrho_{21}=v_{d+1}\dots ,  \varrho_{dd}=v_{d^2}\,.
\]
Then it can be easily verified that the linear representation
on $\mathbb{R}^{d^2}$
 \[
 \boldsymbol{v}^\prime = L\boldsymbol{ v}\,,
\qquad L \in
\mathrm{U}(d)\otimes\overline{\mathrm{U}(d)}\,,
\]
where a line over expression means the complex conjugation,  is isomorphic to the initial adjoint $\mathrm{U}(d)$ action (\ref{eq:adaction}).

Now the corresponding integrity basis ${\cal P} = \left(p_1, p_2, \dots , p_q\right)$ for the ring of invariant polynomials is required  for the  mapping (\ref{eq:polmap}).
For its construction the following observation is in order.   Starting from the center $\mathcal{Z}(\mathfrak{su}(\mathrm{d}))\,$ of the universal enveloping algebra $\mathfrak{U}(\mathfrak{su}(\mathrm{d}))$,   according  to the well-known Gelfand's theorem,
one can define  an isomorphic  commutative symmetrized algebra of invariants
$S(\mathfrak{su}(\mathrm{d}))\,$, which  by turn is isomorphic to the algebra of
invariant polynomials over $\mathfrak{su}(\mathrm{d})$ \cite{Zelobenko}.
The later provides the required resource  for
coordinates that can be used to  parameterize the orbit space $\mathfrak{P}_+/\mathrm{U}(d)\,.$
For our purpose it is convenient to choose the integrity basis that is formed by the so-called trace invariants.
Namely,  we use below the polynomial ring $\mathbb{R}[v_1, v_2,\dots, v_{d^2}]^{\mathrm{U}(d)}=
\mathbb{R}[t_1, t_2, \dots , t_d]\,, $ with n basis elements
\begin{equation}\label{eq:traceinv}
    t_k= \mbox{tr}\left(\varrho^k\right)\,, \qquad k=1,2.\dots, d\,.
\end{equation}
In terms of  the integrity basis  (\ref{eq:traceinv}) the $\mbox{Grad}\--$matrix
reads
\begin{equation}\label{eq:Grad}
\mbox{Grad}(t_1, t_2, \dots , t_d)=\left(
 \begin{array}{ccccc}
 d    & 2t_1     & 3t_2     & \cdots & d t_{d-1} \\
 2t_1  & 2^2t_2     &2\cdot3 t_3    & \cdots &2\cdot d t_{d}   \\
3 t_2  & 2\cdot3t_3     & 3^2t_4    & \cdots & 3\cdot dt_{d+1}   \\
\vdots & \vdots & \vdots & \vdots\,\vdots\,\vdots &\\
d t_{d-1} & 2\cdot d t_{d}&3\cdot d t_{d+1}& \cdots & d^2 t_{2d-2}  \\
 \end{array}
\right)\,.
\end{equation}
In (\ref{eq:Grad}) polynomials $t_k$ with $k>d$ are expressed  as polynomials in $(t_1, t_2, \dots , t_d)$. From (\ref{eq:Grad}) one can easily obtain that
\begin{equation}
\label{eq:Grad2}
\mbox{Grad}(t_1, t_2, \dots , t_d)= \chi \mbox{Disc}\left(
 t_1, t_2, \dots , t_d
\right) \chi^T\,,
\end{equation}
where $\chi=\left( 1,2,\dots, d \right)\,$ and
$\mbox{Disc}\left(
 t_1, t_2, \dots , t_d
\right) $ denotes the matrix
\begin{equation}
\label{eq:DELTA}
\mbox{Disc}
\left(
 t_1, t_2, \dots , t_d
\right)
=
\left(
\begin{array}{ccccc}
 d    & t_1     & t_2     & \cdots &  t_{d-1} \\
 t_1  & t_2     & t_3    & \cdots &t_{d}   \\
 t_2  & t_3     & t_4    & \cdots & t_{d+1}   \\
\vdots & \vdots & \vdots & \vdots\,\vdots\,\vdots &\\
 t_{d-1} & t_{d}& t_{d+1}& \cdots & t_{2d-2}  \\
 \end{array}
\right)\,.
\end{equation}
In one's turn the matrix (\ref{eq:DELTA}) can be written as ``square'' of  the Vandermonde matrix,
\(\mbox{Disc}\left(
 t_1, t_2, \dots , t_d
\right) \ =\Delta\Delta^T\,,
\)
\begin{equation}\label{eq:Vandermode}
 \Delta(x_1,\dots ,x_d) = \left(
 \begin{array}{ccccc}
 1 & x_1 & x_1^2 & \dots &  x_1^{d-1}\\
 1 & x_2 & x_2^2  &\dots & x_2^{d-1}\\
 1 & x_3 & x_3^2  &\dots & x_3^{d-1}\\
 \vdots  & \vdots & \vdots & \vdots\,\vdots\,\vdots &\\
 1 & x_d & x_d^2 &\dots & x_d^{d-1} \\
 \end{array}
 \right)\,,
\end{equation}
whose columns  are determined by powers of roots $(x_1, x_2, \dots x_d)$ of the characteristic equation:
\begin{equation}\label{eq:chareq}
\det|| x - \varrho || =  x^d - S_1x^{d-1} + S_2 x^{d-2}- \dots +(-1)^d\,
S_d  = 0\,.
\end{equation}
The semi-positivity condition of the matrix (\ref{eq:DELTA}) guaranties reality of the roots of (\ref{eq:chareq}). Thus, semi-positivity of the  $\mbox{Grad}\--$matrix is equivalent to the reality condition of eigenvalues of the density  matrix $\varrho$ written in terms of  the $\mathrm{U}(d)$ polynomial   scalars.
Finally, noting that  the density matrices by construction are Hermitian,  we convinced that the Procesi-Schwarz inequalities are satisfied  identically on $\mathfrak{P}_+\,.$

Summarizing,  the algebraic structure of the orbit space $ \mathfrak{P}_+/ \mathrm{U}(d)\, $ is completely  determined
by the inequalities in  elements of the integrity basis for polynomial ring  $
\mathbb{R}[t_1, t_2, \dots , t_d]$ originating from the Hermicity and semi-positivity requirements on density matrices.

\section{Two examples }
\label{sec: examples}

Algebraic structure of the   orbit space of quantum systems is highly intricate. The examples of d=3 (qutrit) and d=4, considered below, demonstrate the degree of its complexity even for the
low dimensional systems.

\subsection{Orbit space of qutrit}
\label{subsec: qutrit}

Qutrit is 3-dimensional quantum system and the integrity basis for $\mathrm{U}(3) $-invariant polynomial consist from
first, second and third order trace polynomials; $t_1, t_2, t_3\,.$ For a visibility below we  consider the case
of normalized density matrices, supposing  $t_1 =1.$\footnote{It is worth to note that  description of the qutrit  orbits is similar to the
studies  of the flavor symmetries of  hadrons,  performed  more than forty ears ago by
by Michel  and Radicati
\cite{MichelRadicati} ( cf.  the method adaptation to  the analysis of space of quantum states \cite{AdelmanCorbettHurnst}, \cite{KusZyczkowski},  \cite{BoyaDixit}).}

\begin{figure}
\begin{center}
  \includegraphics[scale=1.0]{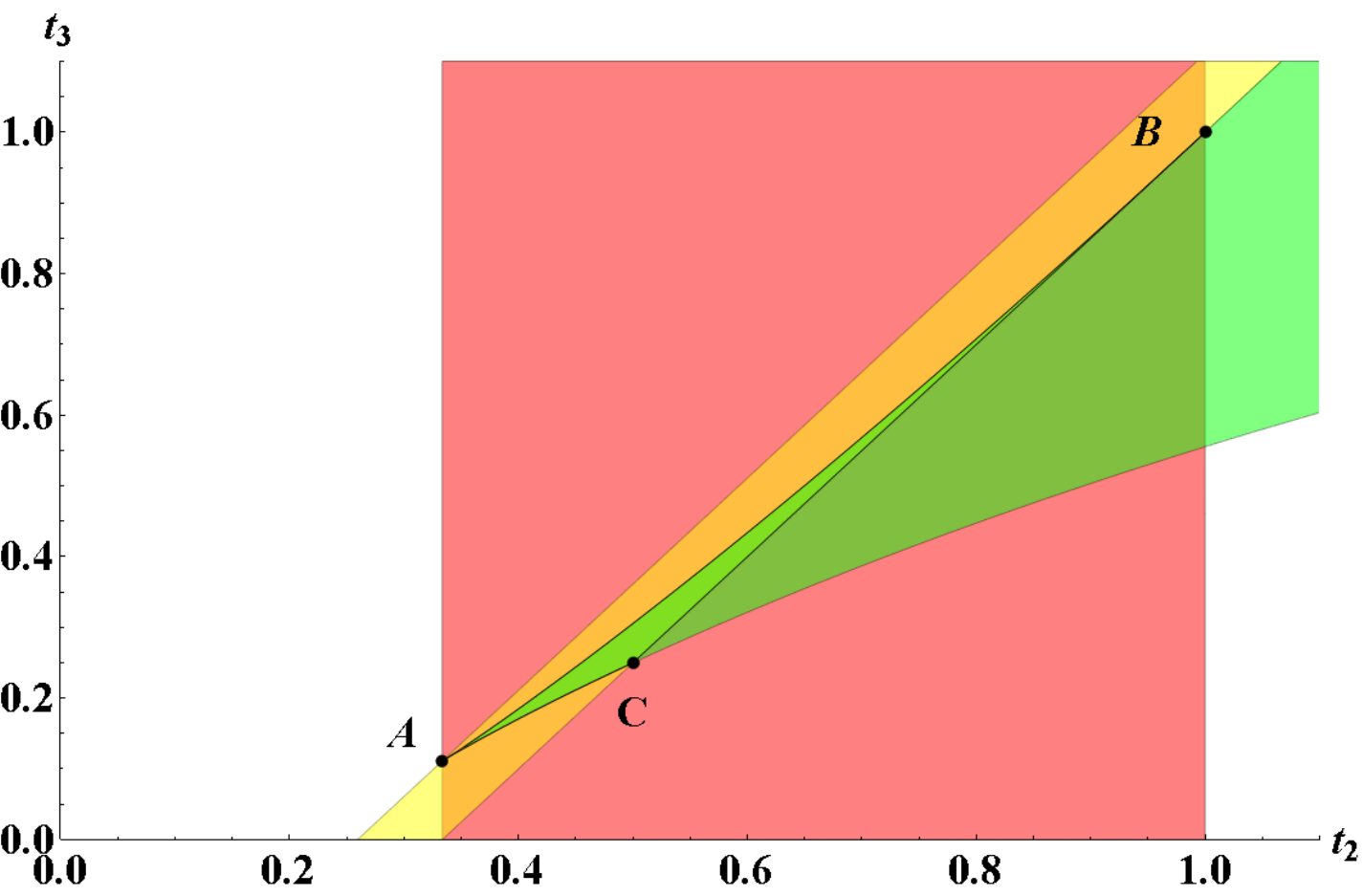}
  \caption{Triangle domain A-B-C  as  the orbit space of  qutrit. }\label{pic. U3T1eq1}
\end{center}
\end{figure}

The condition of  the eigenvalues reality is
\begin{equation}
        0 \leq \frac{1}{6} \left(3 t_2^3-21 t_2^2+36 t_3 t_2+9 t_2-18 t_3^2-8 t_3-1\right)\,,
\end{equation}
while the semi-positivity of density matrices formulated as non-negativity of coefficients of characteristic equation
(\ref{eq:chareq}) reads
\begin{align*}
     & 0\leq\tfrac{1}{2}(1-t_2)\leq \tfrac{1}{3}\,, \\
     & 0\leq\tfrac{1}{2}(1-3t_2+2t_3)\leq \tfrac{1}{9}\,. \\
\end{align*}
Resolving the inequalities
\begin{alignat*}{2}
     & \mbox{Red domain:}       & \qquad   \tfrac{1}{3} \leq & t_2 \leq 1 \\
     & \mbox{Yellow domain: } & \qquad 3t_2-1 \leq 2& t_3 \leq
          3t_2-\tfrac{7}{9} \\
     & \mbox{Green domain:}    &
     \qquad -4+18t_2-\sqrt{2}(3t_2-1)^{3/2} \leq 18 & t_3 \leq -4+18t_2+\sqrt{2}(3t_2-1)^{3/2}
\end{alignat*}
we get the intersection domain shown on Figure \ref{pic. U3T1eq1}.
The triangle domain A-B-C, bounded by the lines:
\begin{alignat*}{2}
    & \mbox{A-B} & \qquad & t_3 =\tfrac{1}{18}( -4+18t_2+\sqrt{2}(3t_2-1)^{3/2}) \\
    & \mbox{A-C} & \qquad & t_3 = \tfrac{1}{18}(-4+18t_2-\sqrt{2}(3t_2-1)^{3/2}) \\
    & \mbox{B-C} & \qquad & t_3=\tfrac{3}{2}t_2-\tfrac{1}{2}
\end{alignat*}
with vertexes \footnote{
Note that the straight line B-C is tangent to the curve A-B at the point $B$
\begin{equation*}
    \frac{dt_3}{dt_2}=1+\tfrac{1}{2\sqrt{2}}(3t_2-1)^{1/2}, \qquad \qquad \frac{dt_3}{dt_2}|_{t_2=1}=\frac{3}{2}\,.
\end{equation*}
}:
$A(\tfrac{1}{3},\tfrac{1}{9})\,,$\
$B(1,1)\,$ and
$ C(\tfrac{1}{2},\tfrac{1}{4})\,,$
represents the orbit space of qutrit in parametrization of trace polynomial coordinates.

Now it is in order to discuss correspondence  between  the above algebraic results and  known classification of orbits with respect to their stability group.
Having in mind this issue consider the Bloch parametrization for qutrit
 \begin{equation}\label{eq:Bloch}
  \rho=\frac{1}{3}\left(\mathbb{I}_3 + \sqrt{3}\,\boldsymbol{\xi}\cdot\boldsymbol{\lambda}\right),
  \end{equation}
where $\boldsymbol{\xi}=(\xi_1, \xi_2, \cdots, \xi_8)\in  \mathbb{R}^8$ denote the Bloch vector  and  $\boldsymbol{\lambda}$  is  the vector, whose components are elements $(\lambda_1, \lambda_2, \cdots, \lambda_8)$ of $\mathfrak{su}(3)$ algebra basis, say the Gell-Mann matrices,
\begin{equation}\label{eq:Gell-Mann}
\begin{array}{c}
\begin{array}{ccc}
 \lambda _{1}=\left(\begin{array}{ccc}
   0 & 1 & 0 \\
   1 & 0 & 0 \\
   0 & 0 & 0 \
 \end{array}\right)&
 \lambda _{2}=\left(\begin{array}{ccc}
   0 & -i & 0 \\
   i & 0 & 0 \\
   0 & 0 & 0 \
 \end{array}\right)&
  \lambda _{3}=\left(\begin{array}{ccc}
   1 & 0 & 0 \\
   0 & -1 & 0 \\
   0 & 0 & 0 \
 \end{array}\right)\\ & & \\
  \lambda _{4}=\left(\begin{array}{ccc}
   0 & 0 & 1 \\
   0 & 0 & 0 \\
   1 & 0 & 0 \
 \end{array}\right)&
 \lambda _{5}=\left(\begin{array}{ccc}
   0 & 0 & -i \\
   0 & 0 & 0 \\
   i & 0 & 0 \
 \end{array}\right)&
  \lambda _{6}=\left(\begin{array}{ccc}
   0 & 0 & 0 \\
   0 & 0 & 1 \\
   0 & 1 & 0 \
 \end{array}\right)\
 \end{array}
 \\ \\
\begin{array}{cc}
\lambda _{7}=\left(\begin{array}{ccc}
   0 & 0 & 0 \\
   0 & 0 & -i \\
   0 & i & 0 \
 \end{array}\right)&
  \lambda _{8}=\displaystyle{\frac{1}{\sqrt{3}}}\left(
  \begin{array}{ccc}
   1 & 0 & 0 \\
   0 & 1 & 0 \\
   0 & 0 & -2 \
\end{array}\right)
\end{array}\
\end{array}\,,
\end{equation}
obeying
\begin{equation}
[\lambda_i , \lambda_j  ] = 2 \imath f_{ijk}\lambda_k
\,,
\qquad \mbox{tr}\left(
\lambda_i\lambda_j
\right)= 2\delta_{ij}\,,
\end{equation}
with non-vanishing structure constants
\begin{equation}\label{eq:strconsu(3)}
f_{123}=2f_{147}=2f_{246}=2f_{257}=2f_{345}=-2f_{156}=-2f_{367}=\frac{2}{\sqrt{3}}f_{458}=\frac{2}{\sqrt{3}}f_{678}=1\,.
\end{equation}
Yo analyse the adjoint orbit
$\cal{O}_\varrho\,$ that passes  through
the point $\varrho$, we define the set of tangent vectors:
\begin{equation}
l_i= \lim_{\theta_1,\theta_2, \dots \theta_8 \to 0}\,
\frac{\partial }{\partial \theta_i}
\left[U\left(
\theta_1,\theta_2, \dots \theta_8\right)\,\varrho\,
U\left(\theta_1,\theta_2, \dots \theta_8
\right)\right]= \imath [\lambda_i, \varrho]\,.
\end{equation}
By definition, the dimension of orbit  $\mbox{dim}(\mathcal{O}_\varrho)$ is given by the
dimension of the tangent space to the orbit $T_{\mathcal{O}_\varrho}$ and therefore equals to the the number of linearly independent vectors  among eight tangent vectors  $l_1, l_2, \dots , l_8\,.$
This number  depends on the point $\varrho$ and
according to the well-known theorem from linear algebra is given by the rank of the so-called  Gram matrix
\begin{equation}\label{eq:GS}
 A_{ij} = \frac{1}{2}\,\Vert \mbox{tr}(l_i l_j)\Vert\,.
\end{equation}
 In  the Bloch parameterization (\ref{eq:Bloch}) we easily find that
 \begin{equation}\label{eq:GSexpresBloch}
 A_{ij} = \frac{4}{3}\,f_{ims}f_{jns}\xi_m\xi_n\,.
\end{equation}

To estimate the rank of matrix (\ref{eq:GS}) it is convenient to pass to the diagonal representative of the matrix $\varrho$:
\begin{equation}\label{eq:diagrep}
\varrho =
W\left(
 \begin{array}{ccc}
 x_1     &  0           & 0 \\
   0       &  x_2      & 0 \\
   0       &  0           & x_3\\
 \end{array}
 \right)W^+\,,
\end{equation}
where $W\in \mathrm{SU(3)}/S_3$ and the descending order for $\varrho$ matrix eigenvalues
\[
1 \geq x_1 \geq x_2 \geq
 x_3 \geq 0\,,
\]
is chosen. The later constraints  allow to avoid a double counting
due to the $S_3\subset  U(3)$ symmetry of permutation of the density  matrix  eigenvalues.
Using the principal  axis transformation (\ref{eq:diagrep}) and taking into account the
adjoint properties  of Gell-Mann matrices
$W^+\lambda_i W = O_{ij}\lambda_j\,,$ with $O \in SO(8)\,,$
the matrix $A_{ij}$ can be written as
\begin{equation}\label{eq:diagonalA}
A_{ij}=O_{ik}A^{\mathrm{diag}}_{kl}O^T_{lj}\,.
\end{equation}
Matrix $A^{\mathrm{diag}}_{kl}$ in  (\ref{eq:diagonalA})  is the matrix (\ref{eq:GS}) constructed from vectors $l^{\mathrm{diag}}_{i}= i [\lambda_i, \varrho_{\mathrm{diag}}]$ tangent to the orbit of  the diagonal matrix
$\varrho_{\mathrm{diag}}:=\mbox{diag}(x_1,x_2, x_3)\,.$

Since we are interesting in determination of $\mbox{rank} \vert A\vert$, the relation
(\ref{eq:diagonalA}) allows to reduce this question to the evaluation of the rank of the diagonal representative $\varrho_{\mathrm{diag}}\,.$
For diagonal matrices the Bloch vector is  $\boldsymbol{\xi}^{\mathrm{diag}}=
(0, 0, 0,  \xi_3, 0, 0, 0, \xi_8)$.  Taking into account the values for structure constants from (\ref{eq:strconsu(3)}), the expression  for $\vert A^{\mathrm{diag}}\vert$ reads
\begin{equation}\label{eq: explidiag}
A^{\mathrm{diag}}=\frac{1}{3}\,\mbox{diag}\left( 4\xi_3^2,  4\xi_3^2,  0,(\xi_3+\sqrt{3}\xi_8)^2,  (\xi_3+\sqrt{3}\xi_8)^2, (\xi_3-\sqrt{3}\xi_8)^2,  (\xi_3-\sqrt{3}\xi_8)^2, 0 \right) \,.
\end{equation}
From (\ref{eq: explidiag}) we conclude that there are orbits of three different dimensions:
\begin{itemize}
\item the orbits of maximal dimension,  $\mbox{dim}(\mathcal{O}_\varrho)=6\,,$
\item the orbits of dimension,  $\mbox{dim}(\mathcal{O}_\varrho)=4\,,$
\item zero dimensional orbit, one point $\boldsymbol{\xi}=0\,.$
\end{itemize}

The above algebraic description  of  the orbits $
\mathcal{O}_\varrho$ corresponds to  their classification  based
on the analysis of the
group of transformations $\mathrm{G}_{\varrho}$\--- the isotropy group (or stability group), which stabilize  point $\varrho \in \mathcal{O}_\varrho$.
The orbits of  different dimensions have a different stability groups; for the points lying on the  orbit of maximal dimension  the stability group is the Cartan subgroup
$\mathrm{U(1)}\otimes \mathrm{U(1)}\otimes\mathrm{U(1)}$, while the stability group of points with diagonal representative $\lambda_8$ is
$\mathrm{U(2)}\otimes \mathrm{U(1)}$.
The dimensions of listed orbits  agrees with the  general formula
\begin{equation}
    \mathrm{dim}\,\mathcal{O}_{\varrho}=
            \mathrm{dim G}-\mathrm{dim {G}_{\varrho}}\,.
\end{equation}
Since the isotropy  group of any two points on the orbit are the same up to conjugation, the orbits can be partitioned into sets with equivalent isotropy groups \footnote{The isotropy group of a point $\varrho$ depends only on the algebraic multiplicity of the eigenvalues of the matrix
$\varrho$.}. This set  is known as {\textit{``strata''}}.

Concluding we refer  to  the relations between the  triangle   A-B-C,
depicted on the  Figure \ref{pic. U3T1eq1},  and the corresponding strata.
The domain inside the triangle ABC corresponds to the principal strata  with the stability group $U(1)\times U(1)\times U(1)$. The discriminant   is positive $|\mbox{Disc}| >0$
and the density matrix  has three different real eigenvalues,
the representative matrix reads  $ \frac{1}{3}(\mathbb{I}_3+\sqrt{3}\left(\xi_3\lambda_3 + \xi_8 \lambda_8\right))$,
with  $\xi_3$ and $\xi_8$ subject to the following constraints
\[
0< 1-\xi_3^2-\xi_8^2 < 1\,,\qquad 0< (2\xi_8-1)(1-\sqrt{3}\xi_3+\xi_8)(1+\sqrt{3}\xi_3+\xi_8) < 1\,.
\]
The $S_3$ coefficient vanishes at line B-C.  The boundary line  B-C, excluding  vertices  B and C also belongs
to the principal stratum, while points B and C belong to the stratum
of lower dimension.   On the sides A-B and A-C the discriminant is zero  $|\mbox{Disc}|=0$, hence, the density matrix  has three real eigenvalues and two of them are equal. At point B two eigenvalues of $\varrho$ are zero.  The  lines  (A-B)/\{A\} and (A-C)/\{A\}  represent the  degenerate  4-dimensional orbits whose stability group is $U(2)\otimes U(1)$. Finally, the  point A is the zero dimensional stratum corresponding to  the maximally mixed state $\varrho= \frac{1}{3} \mathbb{I}_3$.
The details of the orbit types are collected in the Table below.

\begin{equation*}\label{}
    \begin{array}{c|c|c|c|c}
      \dim\mathcal{O} & \mbox{Strata} &
      \mbox{\parbox{3cm}{Stability group}} &
      \mbox{\parbox{4cm}{Representative matrix}}&
      \mbox{\parbox{3.5cm}{Constraints}}
      \\[7mm] \hline\hline\rule{-4pt}{30pt}
      6 & \begin{array}{c}
            \mbox{Interior~of } \\
            \mbox{triangle} \\
            \mbox{ABC}
          \end{array}
       &
        U(1)\otimes
       U(1)\otimes U(1)
      &
      \frac{1}{3}(\mathbb{I}_3+\sqrt{3}\left(\xi_3\lambda_3 + \xi_8 \lambda_8\right))
 & {\mbox{Disc} > 0, S_2>0, S_3>0}
       \\[7mm] \cline{2-5}\rule{-4pt}{30pt}
      & \mbox{\parbox{2.5cm}{Boundary:
      (B-C)/\{B,C\}}}
      &
       U(1)\otimes
       U(1)\otimes U(1)
      &
      \frac{1}{3}(\mathbb{I}_3+\sqrt{3}\left(\xi_3 \lambda_3+ \frac{1}{2}\lambda_8\right))
      & {\mbox{Disc} > 0, S_2>0, S_3=0} \\[7mm] \hline\rule{-4pt}{35pt}
      4 & \begin{array}{c}
      \mbox{Boundary:} \\
            \mbox{(A-B)/\{A\}} \\
            \mbox{(A-C)/\{A\}}
          \end{array}
          &
         U(2)\otimes U(1)
          &
          \begin{array}{c}
          \frac{1}{3}(\mathbb{I}_3+ \sqrt{3}\,\xi_8 \lambda_8)
           \end{array}
      & {\mbox{Disc} = 0, S_2\geq0, S_3\geq0}
      \\[9mm] \hline\rule{-4pt}{30pt}
      0 & \mbox{\parbox{2cm}{Point:
    \{A\}}} & U(3) &
      \frac{1}{3}\mathbb{I}_3
    &  \mbox{Disc}=S_2=S_3=0  \\[9mm] \hline
    \end{array}
\end{equation*}

\vspace{1cm}
\centerline{ Table. The stratum  decomposition for the orbit space of qutrit.}

\subsection{Orbit space of a four-level quantum system}

\begin{figure}
  \includegraphics[width=6cm]{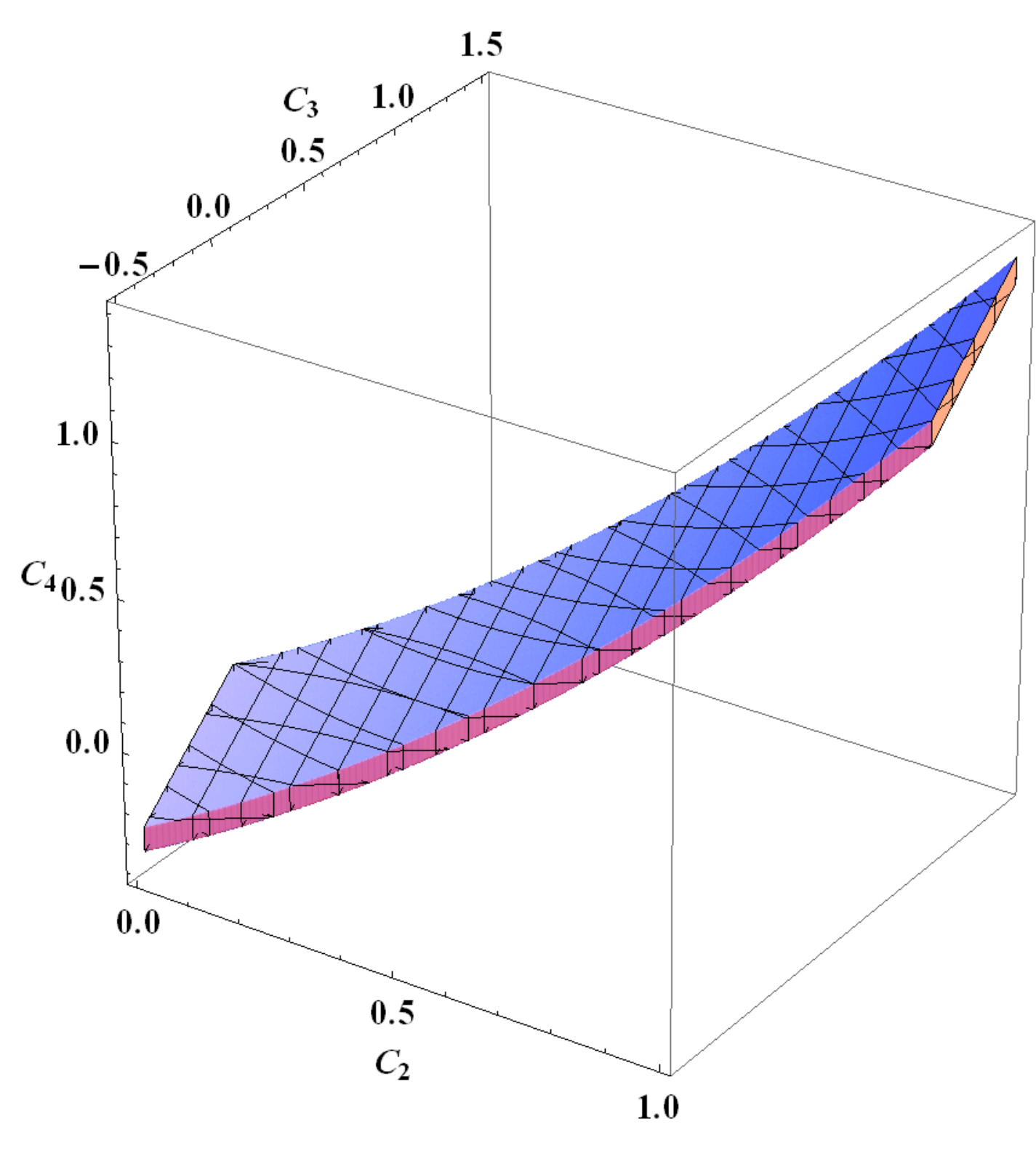}
  \hspace{2cm}
  \includegraphics[width=6cm]{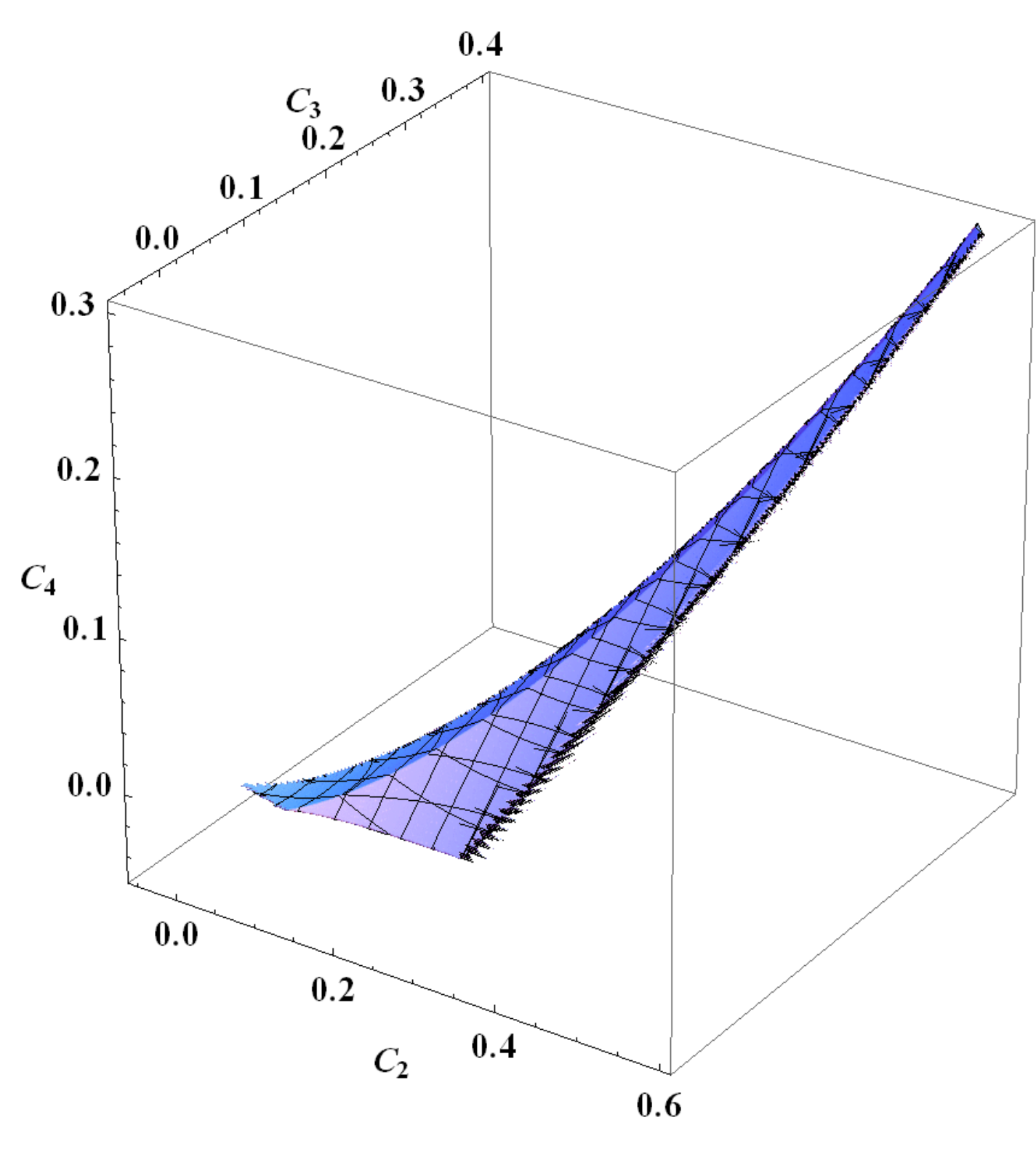}
  \caption{On the left side: $\rho\geq0$;\ On the  right side:
  $\rho \geq 0 \cap  \mbox{Grad}\geq 0$;\label{pic:qudit}}
\end{figure}

The density matrix $\varrho$ of a $4$-level quantum system in the Bloch form reads
\begin{equation} \label{eq:quditBloch}
\rho = \frac{1}{4} \left(\mathbb{I}_4 + \sqrt{6}\,\vec \xi \cdot \vec
\lambda\right)\,,
\end{equation}
where  the traceless part of $\varrho$  is given by scalar product of  15-dimensional
Bloch vector $\vec \xi=\{\xi_1,\ldots,\xi_{15}\} \in
\mathbb{R}^{15}$ with
$\lambda$-vector  whose components are elements of the  Hermitian basis of the Lie algebra
$\mathfrak{su}(4)$
\[
\lambda_i\lambda_j=\frac{1}{2}\,\delta_{ij}\mathbb{I}_4
+(d_{ijk}+i\,f_{ijk})\lambda_k, \qquad i,j,k=1,\ldots,15.
\]
The corresponding integrity basis for the polynomial ring
 $\mathbb{R}[\mathfrak{P}_+]^{\mathrm{U}(4)}\,$ consists of three   $\mathrm{U}(4)$\--invariant polynomials, the Casimir scalars
$\mathfrak{C}_2,\mathfrak{C}_3,\mathfrak{C}_4$
\begin{equation}
\mathfrak{C}_2=\vec{\xi}\cdot\vec{\xi}\,\qquad \mathfrak{C}_3=\sqrt{\frac{3}{2}}\,d_{ijk} \xi_i \xi_j \xi_k\,,
\qquad
\mathfrak{C}_4=\frac{3}{2}\,d_{ijk}d_{lmk} \xi_i \xi_j \xi_l \xi_m\,,
\end{equation}
The semi-positivity of (\ref{eq:quditBloch}) formulated as non-negativity of coefficients
$ S_2, S_3$ and $S_4$ of the characteristic polynomial (\ref{eq:chareq})
\footnote{For details we refer to \cite{GerdtKhvedelidzePalii2009}.}
\begin{eqnarray}\label{eq:quditpos1}
 &S_2=\tfrac{3}{8}(1-\mathfrak{C}_2)\geq 0  \\
  &S_3=\tfrac{1}{16}(1-3\mathfrak{C}_2+2\mathfrak{C}_3)\geq 0,\\
  &S_4=\det\rho=
  \tfrac{1}{256}((1-3\mathfrak{C}_2)^2 +8\mathfrak{C}_3-12\mathfrak{C}_4)\geq 0\label{eq:quditpos3}
\end{eqnarray}
Now we are in position to compute the $\mbox{Grad}$\--matrix in terms of the $\mathrm{SU(4)}$ Casimir scalars:
\begin{equation}\label{eq:Grad4}
\mbox{Grad}=
\begin{pmatrix}
  4\mathfrak{C}_2 & 6\mathfrak{C}_3 & 8\mathfrak{C}_4 \\
  6\mathfrak{C}_3 & 9\mathfrak{C}_4 & 12\mathfrak{C}_2\mathfrak{C}_3 \\
  8\mathfrak{C}_4 & 12\mathfrak{C}_2\mathfrak{C}_3 & 4(\mathfrak{C}_3^2+3\mathfrak{C}_2\mathfrak{C}_4) \\
\end{pmatrix}\,.
\end{equation}
Passing to the equivalent matrix
$Q\,\mbox{Grad}\,Q^T\,,$ with $Q=\mbox{diag}(2, 3, 2 )\,,$ we arrive at
the following form for the Procesi-Schwarz inequalities
\begin{eqnarray}
\label{eq:ProcScw41}
    & \mathfrak{C}_2 + \mathfrak{C}_3^2 + 3 \mathfrak{C}_2 \mathfrak{C}_4 + \mathfrak{C}_4  \geq  0\,,\\
    & \mathfrak{C}_3^2 \left(-4 \mathfrak{C}_2^2+\mathfrak{C}_2+\mathfrak{C}_4-1\right)+\mathfrak{C}_4 \left(3 \mathfrak{C}_2^2+3 \mathfrak{C}_2 \mathfrak{C}_4+\mathfrak{C}_2-4 \mathfrak{C}_4\right) \geq  0\,, \\
    & -4 \mathfrak{C}_2^3 \mathfrak{C}_3^2+3 \mathfrak{C}_2^2 \mathfrak{C}_4^2+6 \mathfrak{C}_2 \mathfrak{C}_3^2 \mathfrak{C}_4-\mathfrak{C}_3^4-4 \mathfrak{C}_4^3 \geq  0\,.
 \label{eq:ProcScw43}
\end{eqnarray}
The domains describing the
semi-positivity of $\rho\,,$ (\ref{eq:quditpos1})-(\ref{eq:quditpos3}), and its residually part  after imposing condition of the semi-positivity of $\mbox{Grad}$\--matrix (\ref{eq:ProcScw41})-(\ref{eq:ProcScw43})
are depicted on the  Figure~\ref{pic:qudit}.

\noindent $\bullet\,${\bf Acknowledgements} $\bullet\,$ The work is supported in part by the Ministry of Education
and Science of the Russian Federation (grant 3802.2012.2) and the Russian Foundation for Basic Research
(grant 13-01-0068).  A. Kh.  acknowledges  the  University of Georgia  for support  under the  grant 07-01-2013.


\end{document}